\documentclass[floatfix,reprint,superscriptaddress,showpacs,preprintnumbers,nofootinbib,amsmath,amssymb,epsfig,pre,floats]{revtex4-1}

\usepackage{amsmath}
\usepackage{amssymb}
\usepackage{epstopdf}
\usepackage{graphicx}
\usepackage{hyperref}
\usepackage{dcolumn}
\usepackage{bm}

\newcommand{\beq}{\begin{equation}}
\newcommand{\eeq}{\end{equation}}
\newcommand{\e}{\mathrm{e}}
\newcommand{\la}{\langle}
\newcommand{\ra}{\rangle}

\PassOptionsToPackage{caption=false}{subfig}
\usepackage{subfig}

\begin{document}

\title{Power-law distribution functions derived from maximum entropy \\ and a symmetry relationship}

\author{G. Jack Peterson}
\affiliation{Biophysics Graduate Group, University of California, San Francisco, CA 94158}
\author{Ken A. Dill}
\email{dill@laufercenter.org}
\affiliation{Laufer Center, and Departments of Physics and Chemistry, Stony Brook University, NY 11794}

\begin{abstract}
Power-law distributions are common, particularly in social physics.  Here, we explore whether power-laws might arise as a consequence of a general variational principle for stochastic processes.  We describe communities of `social particles', where the cost of adding a particle to the community is shared equally between the particle joining the cluster and the particles that are already members of the cluster.  Power-law probability distributions of community sizes arise as a natural consequence of the maximization of entropy, subject to this `equal cost sharing' rule.  We also explore a generalization in which there is unequal sharing of the costs of joining a community.  Distributions change smoothly from exponential to power-law as a function of a sharing-inequality quantity.  This work gives an interpretation of power-law distributions in terms of shared costs.
\end{abstract}


\maketitle

Power-law probability distributions are ubiquitous in nature, especially in social systems.  For example, the fraction $p_k$ of U.S.~cities with a population of $k$ people scales as $p_k \sim k^{-2.37}$~\cite{Zipf_1949,Gabaix_1999,Clauset_2009}.  Other examples of power-law distributions include incomes~\cite{Okuyama_1999}, Internet links~\cite{Broder_2000,Holme_2007}, fluctuations in stock market prices~\cite{Mantegna_1995,Gopikrishnan_1999,Plerou_1999}, company sizes~\cite{Axtell_2001}, numbers of citations received by scientific papers~\cite{Price_1976,Redner_1998,Peterson_2010}, and many others~\cite{Clauset_2009}.

Exponential distributions, such as the Boltzmann distribution law, are also ubiquitous.  But, whereas exponential distributions have a well-known basis in the principles of statistical physics~\cite{Shannon_1948,Jaynes_1957,Shore_1980}, it is unclear if a similar principle underlies power-law distributions.  Here, our interest is in how power-law distributions might arise from stochastic processes, particularly in social physics.  Our approach is based on the principle of maximum entropy (MaxEnt).  MaxEnt is widely used, not only in thermal physics, but also in image analysis~\cite{Wernecke_1977,Gull_1978,Skilling_1984}, drawing inferences~\cite{Jaynes_1979,Skilling_1991}, and in nonequilibrium statistical mechanics~\cite{Stock_2008,Wu_2009,Presse_2011}.  We show here that the same principle, with a `cost-sharing' type of constraint, leads to power-law distributions.

\textbf{Clustering can be framed in terms of `joining costs'.}  We focus on a problem of particle clustering, which provides a convenient language for comparing people joining cities to the statistical physics of growing colloids and polymers.  We first describe a standard growth mechanism, which we express in terms of the `joining costs' of a particle to a growing cluster.  One particle may stick to another, forming a cluster of size 2.  Another particle may join the cluster, forming a size 3 cluster, and so on.  If there are $N$ total particles, then the equilibrium of clusters of various sizes can be written as
\begin{align}\label{eq:schematic}
1 + 1 &\rightleftharpoons 2 \notag\\
1 + 2 &\rightleftharpoons 3 \notag\\
&\vdots\\
1 + (N-1) &\rightleftharpoons N \notag
\end{align}

Let $n_k$ be the number of clusters of size $k$ at equilibrium,
\beq\label{eq:LMA}
n_k = K n_{k-1} n_1 = K^{k-1} n_1 n_2 \cdots n_{k-1},
\eeq
where $K$ is an equilibrium binding constant.  The probability distribution, $p_k$, of cluster sizes is the ratio of the number of size $k$ clusters to the total number of clusters of all sizes,
\beq\label{eq:pXj}
p_k = Q^{-1} n_k,
\eeq
where $Q$ is the grand canonical partition function,
\begin{align}\label{eq:Q}
Q = \sum_{k=1}^N K^{k-1} \prod_{j=1}^{k-1} n_j.
\end{align}

Statistical physics provides an alternative language for expressing the logarithms of populations in terms of energies, free energies or chemical potentials, which are cost-like additive quantities.  Here, we define a dimensionless chemical-potential-like quantity, $\mu_k$, that we call the \emph{joining cost} for a particle to join a size-$k$ cluster,
\beq
\mu_k \equiv \ln \left( K n_k \right).
\eeq
Re-expressing $Q$ in terms of these costs gives
\beq
Q = \sum_{k} \e^{-\sum_{j=1}^{k-1} \mu_j} = \sum_{k} \e^{-w_k},
\eeq
where
\beq\label{eq:genwk}
w_k \equiv \sum_{j=1}^{k-1} \mu_j
\eeq
is the dimensionless cost of assembling the whole cluster.  In this language, if $w_k$ is positive, the distribution will be dominated by small clusters; if $w_k$ is negative, the system will preferentially populate large clusters.  We use this cost language for social physics below.  To proceed further, we need to know how $w_k$ depends on $k$.

\textbf{Independent particle clusters are exponentially distributed.}  First, for illustration, we treat a standard problem of colloid assembly or polymerization.  Assume the cost $\mu_k$ for a particle to join a cluster is independent of $k$.  Then
\beq
\label{eq:indmu}
\mu_{k} = \mu^\circ,
\eeq
where the constant $\mu^\circ$ is the cost of adding one particle to a cluster of any size.  A cluster of size $k$ requires $k-1$ particle additions, so the total cost of assembling the cluster is
\beq
w_k = \mu^\circ \left(k-1\right),
\eeq
and the average cost of adding a particle, taken over all cluster sizes, is
\beq\label{eq:avgWavgk}
\la w \ra \equiv \sum_k w_k p_k = \mu^\circ \left(\la k \ra - 1\right).
\eeq

To predict the probability distribution of cluster sizes, we maximize the entropy,
\beq\label{eq:entropy}
S = -\sum_{k} p_k \ln p_k,
\eeq
subject to two constraints:~(1) a fixed known value of $\la w \ra$ (Eq.~\ref{eq:avgWavgk}) and (2) normalization (ensuring the probabilities sum to 1),
\beq
\sum_{k} p_k = 1.
\eeq
A constraint on $\la w \ra$ gives the extremum function
\beq
\la w \ra - S = \mu^\circ \sum_k (k-1) p_k + \sum_{k} p_k \ln p_k,
\eeq
so that the optimization condition is
\beq\label{eq:indopt}
\sum_{k} dp_k^* \big[ \ln p_k^* + 1 + \alpha + \mu^\circ\left(k-1\right) \big] = 0,
\eeq
where $\alpha$ and $\mu^\circ$ are the Lagrange multipliers that enforce normalization and constraint~\ref{eq:avgWavgk}, respectively.  Solving Eq.~\ref{eq:indopt} gives the equilibrium probability distribution, $p_k^*$, which maximizes the entropy and satisfies the average cost and normalization constraints.  The solution is the standard exponential distribution of cluster sizes, 
\beq\label{eq:pjexp}
p_k^* = \e^{-1-\alpha} \e^{-\mu^\circ\left(k-1\right)} = Q^{-1}\, \e^{-\mu^\circ k},
\eeq
where $Q$ is the grand canonical partition function,
\beq\label{eq:QRA}
Q = \sum_{k=1}^N \e^{-\mu^\circ k} = \frac{1-\e^{-\mu^\circ N}}{1-\e^{-\mu^\circ}}.
\eeq

\textbf{Equal cost sharing leads to power-law distributions.}  Now, we consider a more general notion of a particle's joining cost when it enters a $k$-mer cluster.  A person is more likely to join a larger city than a smaller city because of greater opportunities of jobs, infrastructure, services, entertainment, and other economic and social factors.  Existing citizens have already paid some of the cost of entry for the new joiner `particle'.  So, relative to the cost of joining an independent-particle cluster ($\mu^\circ$), the cost $\mu_k$ of joining a `social-particle' cluster of size $k$ is reduced to
\beq\label{eq:reduction}
\mu_k = \mu^\circ - k r_k.
\eeq
Eq.~\ref{eq:reduction} expresses the idea of \emph{cost sharing}, namely that the cost of joining a cluster is reduced because the existing $k$ member particles provide a discount of $r_k$ each to the joiner particle.

There are two non-arbitrary limiting cases for how we might choose the value of $r_k$: \textbf{(1) No sharing}, $r_k = 0$, and the particles are independent, as described above, or \textbf{(2) Equal sharing}, where each member pays the same amount as the joiner when it enters the cluster,
\beq\label{eq:rkmuk}
r_k = \mu_k.
\eeq
Substituting Eq.~\ref{eq:rkmuk} into \ref{eq:reduction} and solving for $\mu_k$ yields
\beq
\label{eq:widget}
\mu_k = \frac{\mu^\circ}{1+k},
\eeq
which expresses how the costs diminish with cluster size in social-particle systems.  The total cost to assemble a social cluster of $k$ particles is
\beq
w_k = \mu^\circ \sum_{j=1}^{k-1} \frac{1}{1+j},
\eeq
which can be expressed as (see Appendix)
\beq\label{eq:Eklog}
w_k \approx \mu^\circ\left[\ln \left(k+\frac{1}{2}\right) + \gamma - 1\right],
\eeq
where $\gamma = 0.5772...$ is Euler's constant.  The average joining cost per particle is
\beq\label{eq:ECSconstr}
\la w \ra \approx \mu^\circ \left[ \left\la \ln \left(k+\frac{1}{2}\right) \right\ra + \gamma -1 \right].
\eeq

To predict the social-particle probability distribution, we maximize the entropy subject to constraint~\ref{eq:ECSconstr} on $\la w \ra$,
\beq
\sum_k dp_k^* \left\{ \ln p_k^* + 1 + \alpha + \mu^\circ \left[ \ln \left(k+\frac{1}{2}\right) + \gamma -1 \right] \right\} \approx 0.
\eeq
Solving for $p_k^*$ yields
\beq\label{eq:pkstirling}
p_k^* \sim \left(k+\frac{1}{2}\right)^{-\mu^\circ},
\eeq
giving a power-law distribution.  The scaling exponent $\mu^\circ$ is the `un-discounted cost' of adding one particle to a cluster.\footnote{This scaling form for $p_k^*$ (\ref{eq:pkstirling}) is relatively accurate for most values of $k$ ($k \gtrsim 2$).  The exact forms, valid for all values of $k$, are given in the Appendix.}

The present work shows how power-law distributions can emerge naturally from random clustering of particles that equally share the joining cost.  The `rich-get-richer' aspect of power-law size distributions is that social particles are more attracted to bigger clusters than to smaller clusters.  We have expressed this in a language of `costs': the power-law arises because member particles equally share the joining costs with joiner particles.  The power-law exponent is the `un-reduced' cost $\mu^\circ$.\footnote{Our term `costs' here can alternatively be thought of in terms of `economies of scale'.  Making more widgets decreases the cost-per-widget.  That is, if the cost of a widget factory is $\mu^\circ$, then the cost per widget for the first widget is $\mu^\circ$ and the cost per widget for making two widgets is $\mu^\circ/2$.  In this sense, the cost of the second is `shared' by the first.}

The aim of the present work is not to build a model of a specific power-law process.  We do not give here a generative way to derive $\mu^\circ$ for any particular problem of social physics.  Rather, the present treatment plays a role more like the Boltzmann law, which is a framework for models, rather than a particular model itself.  There is a large literature of models that generate power laws~\cite{Mitzenmacher_2004,Newman_2005}.  We believe that, at least in some cases, such models may be interpretable in terms of shared costs.

In many situations involving people, the idea of cost sharing arises naturally.  For example, a well-studied power-law distribution is of the number of citations to scientific papers~\cite{Price_1976,Redner_1998,Peterson_2010}.  Suppose the author of paper A is preparing to cite paper C.  Every paper in the set \{B\} that has already cited paper C can be thought of as a member of a community that paper A is about to join.  The larger is the community \{B\}, the more likely it is for paper A to join that community, and cite paper C.  The community \{B\} has already paid a higher cost by finding paper C in sea of options that was larger at the time.  In this way, any paper $B$ has lowered the cost for the author of paper $A$ to find and cite paper C.

The distribution of U.S.~city sizes also has a power-law tail~\cite{Zipf_1949,Gabaix_1999}.  It is more likely a person will choose to move to Los Angeles, CA (population: 9 million) than to Fields, Oregon (population: 86).  Similar to the cost amortization obtained through an economy of scale, the people of LA have already paid the development costs of creating the companies, jobs, infrastructure, and services that attract new individuals.  Thus, the marginal cost of adding one more person is reduced, relative to the cost of adding a person to a smaller city.

\textbf{A generalized model for partial cost sharing.}  Described above are two limiting cases: no cost sharing (independent particles) or full cost sharing (the members pay the same as the joiner).  What if the member particles only pay a fraction $s$ of the cost that the joiner particle pays?  Now the cost reduction is
\beq\label{eq:unequal}
r_k = s\mu_k.
\eeq
Combining Eq.~\ref{eq:unequal} with Eq.~\ref{eq:reduction} and solving for $\mu_k$, we find
\beq
\mu_k = \frac{\mu^\circ}{1+sk},
\eeq
so the total cost of assembling a partially-social cluster of size $k$ is
\beq\label{eq:Wkh}
w_k = \mu^\circ \sum_{j=1}^{k-1} \frac{1}{1+sj} \approx \frac{\mu^\circ}{s} \ln \left(k+ \frac{1}{s}-\frac{1}{2} \right) + \text{const},
\eeq
where the constant will be absorbed into the normalization (see Appendix).  Maximizing the entropy gives
\beq\label{eq:mixedpk}
p_k^* \sim \left( k+ \frac{1}{s}-\frac{1}{2}\right)^{-\mu^\circ/s}.
\eeq

\begin{figure}
\centering{
\includegraphics[width=0.48\textwidth]{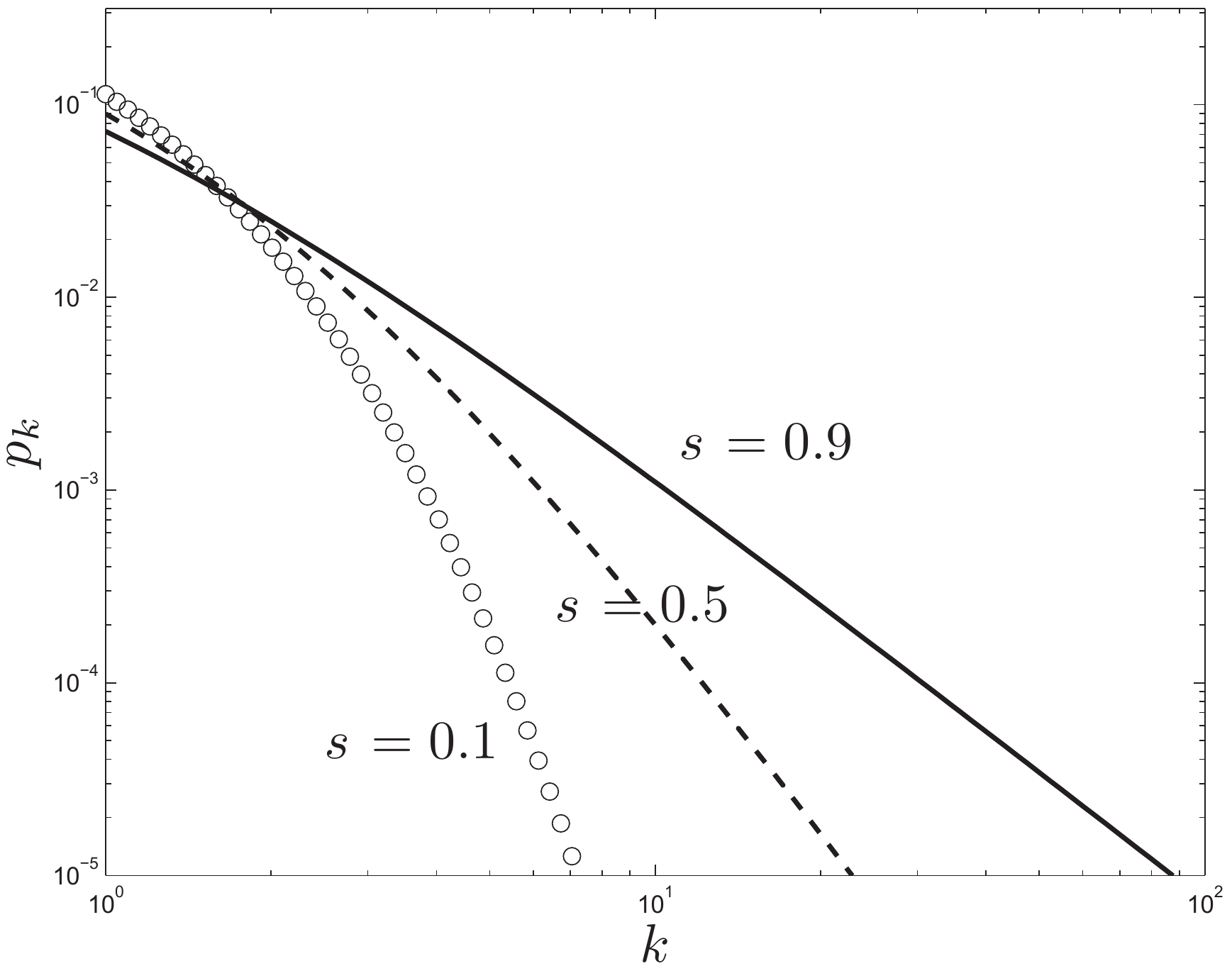}
}
\caption{Probability distributions for various sharing values, with $\mu^\circ = 2$.  At $s=0.1$, the particles are barely helpful at all, and the entering particle must pay most of its joining cost, so the distribution is nearly exponential.  At $s=0.9$, the particles are very helpful, and the distribution is a power-law.}
\label{fig:mixedpk}
\end{figure}

We call $s$ the sharing parameter: $s=0$ involves no sharing (\emph{i.e.}, independent particles) and $s=1$ involves full equal sharing between the joiner particle and all members.  The $s$ parameter controls the shape of the distribution.  For $s=0$, the entry cost is the same for all clusters, and the distribution is exponential; there are very few large clusters.  For $s=1$, the particles are social, and the entry cost is reduced for larger clusters, resulting in a power-law distribution; there are more large clusters in this case.  Fig.~\ref{fig:mixedpk} shows that varying $s$ changes the distribution smoothly from exponential to power-law.  We refer to the limit $s \rightarrow \infty$ as `super-social particles': the existing members pay the full cost, and the joiner particle pays nothing.  For super-social particles, we obtain a uniform distribution; all cluster sizes are equally probable.

We have shown how power-law distributions can arise naturally from the maximization of entropy subject to a symmetry relationship in which all particles share the cost incurred when a new particle joins a cluster.  It has been noted before that MaxEnt with logarithmic constraints leads to power-law distributions \cite{Milakovic_2001,Harte_2011}.\footnote{Logarithmic constraints have also been justified in terms of the information content of language~\cite{Mandelbrot_1953}, to explain the observation that word-use frequency obeys a power-law~\cite{Zipf_1949,Newman_2005}.}  Exponential distributions result from constraints on linear averages such as $\langle k \rangle$, while power-law distributions result from constraints on logarithmic averages such as $\langle \ln k \rangle$.  Here, we have described how such constraints can be interpreted as a type symmetry of sharing that is natural in the social realm.

Many previous studies on generative mechanisms for power-laws have investigated a family of `proportional attachment' (PA) rules~\cite{Yule_1925,Simon_1955,Price_1976,Barabasi_1999}.  In the context of particle clustering, a PA rule says that the probability of a cluster acquiring a new particle is proportional to the number of particles it already contains.  A premise of the PA rule is that the joining particle is cognizant of the populations of the clusters in the system.  By contrast, our cost-sharing framework does not assume the particles know anything about the system, so the present approach may be useful for modeling systems composed of `uninformed' particles.

We thank H.~Ge, S.~Press\'{e}, and C.~Shalizi for helpful discussions.  GJP thanks the Department of Defense for financial support from a National Defense Science and Engineering Graduate Fellowship.  KAD thanks the National Science Foundation and National Institutes of Health GM 34993.

\bibliographystyle{unsrt}

\cleardoublepage

\begin{appendix}
\section*{Appendix}

\textbf{Approximate cost function for social particles.}  Here, we derive expression \ref{eq:Eklog} by including order $1/k$ corrections from the sum-to-integral conversion of $w_k$ for social particles.  We use the Euler-Maclaurin formula \cite{Abramowitz_1972} to convert the sum in Eq.~\ref{eq:widget} to an integral:
\beq\label{eq:EM}
\sum_{j=1}^{k-1} \frac{1}{1+j} = \ln k + \frac{1}{2k} + C + \mathcal{O}\left(k^{-2}\right),
\eeq
where $C$ is an unknown constant.

First, we calculate the asymptotic form of $\ln (k+a)$ (for some constant $a$, where $k \gg a$) by Taylor expanding around $a=0$:
\beq\label{eq:logasympt}
\ln (k+a) \sim \ln k + \frac{a}{k} + \mathcal{O}\left(k^{-2}\right).
\eeq
We use \ref{eq:logasympt} to absorb the $1/k$ term in \ref{eq:EM} into the logarithm,
\beq\label{eq:absorbed}
\ln k + \frac{1}{2k} \sim \ln \left( k + \frac{1}{2} \right).
\eeq
Next, we evaluate the constant $C$.  We begin by defining the $k^\text{th}$ harmonic number,
\beq
H_k \equiv \sum_{j=1}^k \frac{1}{j} = \psi \left(1+k\right) + \gamma,
\eeq
where $\psi (k) \equiv d\ln \Gamma(k)/dk$ is the digamma function, and $\gamma = 0.5772...$ is Euler's constant.  The sum in Eq.~\ref{eq:widget} can be written
\beq\label{eq:ECSharmonic}
\sum_{j=1}^{k-1} \frac{1}{1+j} = H_k  - 1 = \psi \left(1+k\right) + \gamma - 1.
\eeq
Ignoring the term in Eq.~\ref{eq:ECSharmonic} that depends on $k$, we see that the remaining constant terms are
\beq\label{eq:c}
C = \gamma - 1.
\eeq
Substituting \ref{eq:absorbed} and \ref{eq:c} into \ref{eq:EM}, we find
\beq
\sum_{j=1}^{k-1} \frac{1}{1+j} \sim \ln\left(k+\frac{1}{2}\right) + \gamma - 1 + \mathcal{O}\left(k^{-2}\right).
\eeq
Dropping the order $1/k^2$ corrections gives the expression used in \ref{eq:Eklog}.

\textbf{Exact distributions.}  In Eq.~\ref{eq:Eklog}, we have given an approximation to $p_k$.  Here, we use Eq.~\ref{eq:ECSharmonic} to obtain the exact expression for the probability distribution function $p_k$.  We write the cost function $w_k$ as
\beq\label{eq:exactsocialwk}
w_k = \mu^\circ \sum_{j=1}^{k-1} \frac{1}{1+j} = \mu^\circ \Big[\psi \left(1+k\right) + \gamma -1\Big].
\eeq
Maximizing the entropy yields
\beq\label{eq:exactsocialpk}
p_k^* = Q^{-1} \e^{-\mu^\circ \psi \left(1+k\right) },
\eeq
with partition function
\begin{align}
Q &= \sum_{k=1}^N \e^{-\mu^\circ \psi \left(1+k\right) }.
\end{align}

In the more general case of partially social particles, the cost function is
\beq\label{eq:partialsocialwk}
w_k = \mu^\circ \sum_{j=1}^{k-1} \frac{1}{1+sj} = \frac{\mu^\circ}{s} \left[\psi \left(\frac{1}{s}+k\right) - \psi\left(1+\frac{1}{s} \right)\right],
\eeq
leading to the probability distribution
\beq\label{eq:partialsocialpk}
p_k^* = Q^{-1} \e^{-\frac{\mu^\circ}{s} \psi \left(\frac{1}{s}+k\right) },
\eeq
where
\beq
Q = {\sum_{k=1}^N \e^{-\frac{\mu^\circ}{s} \psi \left(\frac{1}{s}+k\right)}}.
\eeq

\end{appendix}

\end{document}